\documentclass{article}
\usepackage{emulateapj}

\def\lsun{{\rm L_{\odot}}}
\def\msun{{\rm M_{\odot}}}
\def\rsun{{\rm R_{\odot}}}

\def\be{\begin{equation}}
\def\ee{\end{equation}}
\begin{document}

\title{RADIATIVELY--DRIVEN OUTFLOWS AND AVOIDANCE OF COMMON--ENVELOPE 
EVOLUTION IN CLOSE BINARIES }

\author{Andrew~R.~King\altaffilmark{1, 2} and Mitchell~C.~Begelman\altaffilmark{1, 3}}
\authoremail{ark@star.le.ac.uk, mitch@jila.colorado.edu}

\altaffiltext{1} {Institute for Theoretical Physics, University of
California at Santa Barbara, CA 93106-4030, U.S.A.}

\altaffiltext{2} {Astronomy Group, University of Leicester,
Leicester LE1 7RH, U.K.}

\altaffiltext{3}{JILA, University of Colorado, Boulder, CO 80309-0440,
U.S.A.}

\begin{abstract}

Recent work on Cygnus X--2 strongly suggests that neutron--star or
black hole binaries survive highly super--Eddington mass transfer
rates without undergoing common--envelope evolution. We suggest here
that the accretion flows in such cases are radiation--pressure
dominated versions of the `ADIOS' picture proposed by Blandford \&
Begelman (1999), in which almost all the mass is expelled from large
radii in the accretion disc. We estimate the maximum radius from which mass loss is likely to occur, 
and show that common--envelope evolution is probably avoided in any binary in
which a main--sequence donor transfers mass on a thermal timescale to
a neutron star or black hole, even though the mass transfer rate may
reach values $\sim 10^{-3}~\msun\ {\rm yr}^{-1}$. This conclusion
probably applies also to donors expanding across the Hertzsprung gap,
provided that their envelopes are radiative. SS433 may be an example
of a system in this state.

\end{abstract}


\section{INTRODUCTION}

The question of what happens to a compact object that is fed mass at
rates far higher than its Eddington limit has a long history (Shakura
\& Sunyaev, 1973; Kafka \& M\'esz\'aros, 1976; Begelman, 1979).  In
the context of accreting binary systems, this problem is particularly
acute because of the possibility of common--envelope (CE) evolution at
such rates. That is, the accreting component may be unable either to
accept or to expel the mass at a sufficiently high rate to avoid the
formation of an envelope engulfing the entire binary system. The
frictional drag of this envelope can shrink the binary orbit
drastically. If the resulting release of orbital energy is enough to
unbind the envelope, the binary will emerge from the common envelope
with a smaller separation; if not, the binary components may
coalesce. CE evolution is probably required for the formation of
binaries such as cataclysmic variables, in which the binary separation
is far smaller than the radius of the accreting white dwarf's red
giant progenitor. However, it is in general an open question whether CE
evolution occurs in any given binary.

This question is thrown into sharp relief by recent work on the
evolution of the low--mass X--ray binary Cygnus X--2 (King \& Ritter,
1999), which has a period of 9.84~d. The rather precise spectroscopic
information found by Casares, Charles, \& Kuulkers (1998), together with the observed
effective temperature of the secondary, shows that this star has a
mass definitely below $0.7\msun$ and yet a luminosity of order
$150\lsun$. King \& Ritter (1999) consider several possible
explanations and show that the only viable one is that Cygnus X--2 is
a product of early massive Case B evolution. Here `Case B' means that
the mass--losing star has finished core hydrogen--burning, and is
expanding across the Hertzsprung gap: `early' means that the stellar
envelope is radiative rather than convective, and `massive' that the
helium core is non--degenerate; see Kippenhahn \& Weigert, 1967. In
Cygnus X--2 an initially more massive ($M_{2i} \simeq 3.5\msun$)
secondary transferred mass on a thermal timescale ($\sim 10^6$~yr) to
the neutron star. This idea gives a satisfying fit to the present
observed properties of Cyg X--2, as well as a natural explanation for
the large white dwarf companion masses found in several millisecond
pulsar binaries with short orbital periods. CE evolution cannot have
occurred, as Cyg X--2's long orbital period means that there was far
too little orbital energy available for the CE mechanism to have
ejected so much mass. Thus an inescapable feature of this picture is
that the neutron star is evidently able to eject essentially all of
the matter ($\ga 2 - 3\msun$) transferred to it at highly super--Eddington
rates $\ga 10^{-6}\msun$~yr$^{-1}$.  Indeed, the neutron star mass in
Cyg X--2 is rather close to the canonical value of $1.4\msun$. The aim
of this paper is to determine under what conditions such expulsion can
occur without the system going into a common envelope.

\section{EXPULSION BY RADIATION PRESSURE}

There are essentially two views as to the fate of matter dumped onto a
compact object at a highly super--Eddington rate. In
spherically--symmetric, dissipative accretion of an
electron--scattering medium, the luminosity generated by infall down
to radius $R$ will reach the Eddington limit at a radius 
\be 
R_{\rm ex} \sim \biggl({\dot M_{\rm tr}\over \dot M_{\rm Edd}}\biggr)R_S, 
\ee 
where $\dot M_{\rm tr}$ is the mass infall rate at large radius (i.e. the mass
transfer rate from the companion star in our case), $\dot M_{\rm Edd} =
L_{\rm Edd}/ c^2$ is
the Eddington accretion rate, and $R_S$ is the Schwarzschild radius
(Begelman, 1979). This is also the ``trapping radius", below which
photon diffusion outward cannot overcome the advection of photons
inward.  If the compact object is a black hole, the radiation generated
in excess of the Eddington limit can thus be swept into the black hole,
and lost.  If the compact object is a neutron star, however, radiation
pressure building up near the star's surface must resist inflow in
excess of $\dot M_{\rm Edd}$, causing the stalled envelope to grow
outward.  This situation would lead to the formation of a common
envelope.

The outcome may be very different if the accretion flow has even a
small amount of angular momentum.  Shakura \& Sunyaev (1973) suggested
that super--Eddington flow in an accretion disk would lead to the
formation of a strong wind perpendicular to the disk surface, which
could carry away most of the mass. Such a model
(an ``Adiabatic Inflow-Outflow Solution,", or ADIOS) was
elaborated by Blandford \& Begelman (1999: hereafter BB99), who
considered radiatively inefficient accretion flows in general.  BB99
recalled that viscous transfer of angular momentum also entails the
transfer of energy outward. If the disk were unable to radiate
efficiently (as would be the case at $R < R_{\rm tr}$), the energy
deposited in the material well away from the inner boundary would
unbind it, leading to the creation of powerful wind.  BB99 described a
family of self-similar models in which the mass inflow rate decreases
inward as $\dot M \propto r^n$ with $0<n<1$.  The exact value of $n$
depends on the physical processes depositing energy and
angular momentum in the wind.  If these are very efficient (e.g.,
mediated by highly organized magnetic torques) $n$ could be close to
zero, in which case little mass would be lost.  However, if the wind
is produced inefficiently, $n$ would have to be close to 1 and the
mass flux reaching the central parts of the accretion disk would be
much smaller than the mass transferred from the secondary.  For
example, two-dimensional hydrodynamical simulations of the evolution
of a non-radiative viscous torus (Stone, Pringle \& Begelman 1999)
show the development of a convectively driven circulation with little
mass reaching the central object, and $n\sim 1$.  The development of this
strong mass loss is generic and is not related to the assumption of
self-similarity. In effect, what is happening is that the energy
liberated by a small fraction of the mass reaching the deep
gravitional potential serves to unbind the majority of the matter
which is weakly bound at large distances.

While the specific details of mass loss from super--Eddington flow have
not been worked out (in particular, radiation-dominated convection is
poorly understood), it is reasonable to assume that the wind will be
produced inefficiently, with $n\sim 1$ as in the case of hydrodynamic
convection. We also assume that most of the matter will be blown away
from $R_{\rm ex}$.  Applying equation (1), we find
\be
R_{\rm ex} \simeq 1.3\times 10^{14}\dot m_{\rm tr} \ {\rm cm},
\label{rex2}
\ee
where $\dot m_{\rm tr}$ is the mass transfer rate expressed in $\msun$
yr$^{-1}$.  Note that $R_{\rm ex}$ is independent of the mass of the
compact accretor. Since we restrict attention to electron scattering
opacity only, we require that hydrogen should be strongly ionized at
$R_{\rm ex}$. This is ensured by requiring the radiation temperature
near $R_{\rm ex}$ to
exceed $T_H \sim 10^4$~K. Since the luminosity emerging from $R_{\rm
ex}$ is close to the Eddington limit, we require
$L_{\rm Edd} \ga 4\pi R_{\rm ex}^2\sigma T_H^4$, which is satisfied if
$R_{\rm ex}/R_S \la 10^7 m_1^{-1/2}$, or equivalently (using
equation[1])
\be
\dot M_{\rm tr}\la  10^7 \dot M_{\rm Edd} m_1^{-1/2}
      \simeq 2\times 10^{-2} m_1^{1/2} ~\msun\ {\rm yr}^{-1}
\label{h}
\ee
where $m_1 = M_1/\msun$ is the mass of the compact accretor (black hole
or neutron
star).

CE evolution will be avoided if $R_{\rm ex}$ is smaller than the
accretor's Roche lobe radius $R_1$. If the accretor is the less
massive star (as will generally hold in cases of interest) we can use
standard formulae to write
\be
r_1 = 1.9m_1^{1/3}P_{\rm d}^{2/3},
\label{roche}
\ee
where $r_1 = R_1/\rsun$ and $P_{\rm d}$ is the orbital period measured
in
days. Combining with equation (\ref{rex2}) gives
\be
\dot M_{\rm tr} \la
10^{-3}m_1^{1/3}P_{\rm d}^{2/3}~\msun\ \ {\rm yr}^{-1}.
\label{lim1}
\ee
This form of the limit can be compared directly with observation if we
have estimates of the transfer rate, orbital period and the accretor
mass.
For more systematic study it is useful to replace the dependence on the
accretor's Roche lobe by that on its companion's. Thus,
since the mass transfer rate
is specified by properties of the companion star, which is assumed
to fill its Roche lobe radius $R_2$, we eliminate $R_1$ from the
condition
$R_{\rm ex} \la R_1$ by using the relation
\be
{R_1\over R_2} = \biggl({m_1\over m_2}\biggr)^{0.45},
\label{lobe}
\ee
(cf King et al., 1997)
where $M_2 = m_2\msun$ is the companion mass. Writing $R_2 = r_2\rsun$
we
finally get the limit
\be
\dot M_{\rm tr} \la
5\times 10^{-4}m_1^{0.45}m_2^{-0.45}r_2~\msun\ {\rm yr}^{-1}
\label{lim}
\ee
on the mass transfer rate if CE evolution is not to occur.

\section{AVOIDANCE OF COMMON ENVELOPE EVOLUTION}

By specifying the nature of the companion star we fix $m_2, r_2$ and
$\dot M_{\rm tr}$ in (\ref{lim}), and so can examine whether CE
evolution is likely in any given case. Rapid mass transfer occurs if
the companion star is rather more massive than the accretor, since
then the act of transferring mass shrinks the donor's Roche lobe. The
mass transfer proceeds on a dynamical or thermal timescale depending
on whether the donor star's envelope is largely convective or
radiative (e.g.  Savonije, 1983). In the first case, CE evolution is
quite likely to ensue, as the mass transfer rate rises to very high
values. However, even in this case it is worth checking the inequality
(\ref{lim}) in numerical calculations, as the e--folding time for the
mass transfer is $t_e\sim (H/R_2)t_M$, where $H$ is the stellar
scaleheight and $t_M$ is the mass transfer timescale set by whatever
process (e.g., nuclear evolution) brought the donor into contact with
its Roche lobe initially. For main--sequence and evolved stars we have
$H/R_2 \sim 10^{-4}, 10^{-2}$ respectively. Thus $t_e$ may be long
enough that the companion mass is exhausted before (\ref{lim}) is
violated.

Thermal--timescale mass transfer is rather gentler, and offers the
possibility of avoidance of CE evolution. In addition to the case
mentioned above, thermal--timescale mass transfer will also occur if
the donor star is crossing the Hertzsprung gap and has not yet
developed a convective envelope (i.e., is not close to the Hayashi
line), even if it is the less massive star. Detailed calculations
(Kolb, 1998) show that in both cases the mass transfer rate is given
roughly by
\be
\dot M_{\rm tr} \sim {M_2\over t_{\rm KH}},
\label{th}
\ee
where
\be
t_{\rm KH} = 3\times 10^7{m_2^2\over r_2l_2}~{\rm yr}
\label{kh}
\ee
was the Kelvin--Helmholtz time of the star when it left the main
sequence,
and $L_2 = l_2\lsun$ was its luminosity. (Note that by definition the donor
is not in thermal equilibrium, so an originally main--sequence donor
will develop a non--equilibrium structure as mass transfer proceeds.)
The condition of a radiative envelope requires a main--sequence mass 
$m_2 \ga 1$, so we may take
\be
r_2 \sim m_2^{0.8},\ \  l_2 \sim m_2^3.
\label{ms}
\ee
Inserting in (\ref{kh}) and (\ref{th}) we find
\be
\dot M_{\rm tr} \sim 3\times 10^{-8}m_2^{2.8},
\label{trmax}
\ee
so comparing with (\ref{lim}) we require
\be
m_2 \la 53 m_1^{0.18}
\label{mmax}
\ee
and thus (from \ref{trmax})
\be
\dot M_{\rm tr,\ max} \sim 2\times 10^{-3}m_1^{0.51}\ \msun\ \ {\rm yr}^{-1}.
\label{mtrmax}
\ee
Hence we expect CE evolution to be avoided in thermal--timescale mass
transfer from a main--sequence star,
or from a Hertzsprung gap star, provided that it
has a radiative envelope. This is
in agreement with the assumption of no CE evolution
in Cyg X--2 made by King \& Ritter (1999), where the initial donor mass
was about $3.5\msun$.

\section{CONCLUSIONS}

We have derived a general criterion for the avoidance of
common--envelope evolution in a binary in which the accretor is a
neutron star or a black hole. This shows that thermal--timescale mass
transfer from a main--sequence star is unlikely to lead to CE
evolution, as is mass transfer from a Hertzsprung gap star, provided
that the envelope is radiative.  The first possibility allows the
early massive Case B evolution inferred by King \& Ritter (1999) for
the progenitor of Cyg X--2.  SS433 may be an example of the second
possibility, with a fairly massive donor star. We will discuss this
possibility in detail in a future paper.

The considerations of this paper suggest that common--envelope
evolution with a neutron--star or black--hole accretor generally
requires an evolved donor with a deep convective envelope. This
represents a slight restriction on some of the routes invoked in the
possible formation of Thorne--$\dot {\rm Z}$ytkow objects.

\acknowledgements

This research was carried out at the Institute for Theoretical Physics
and supported in part by the National Science Foundation under Grant
No. PHY94--07194. ARK gratefully acknowledges support by the UK
Particle Physics and Astronomy Research Council through a Senior
Fellowship. MCB acknowledges support from NSF grant AST95--29170 and a
Guggenheim Fellowship.

%

\end{document}